\documentclass[11pt]{article}
\usepackage{epsfig,amsbsy,calc,graphics,latexsym,graphicx,psfrag,color,citesort}

\newcommand{\nc}{\newcommand}
\newcommand{\nt}{\newtheorem}

\newcounter{Diagrams}
\Alph{Diagrams}
\newcounter{cancellation}
\addtocounter{cancellation}{1}

\nt{D}{}[Diagrams]
\nt{DC}[D]{\{}
\nt{cancel}{Cancellation}

\nc{\QED}{QED}
\nc{\ERG}{ERG}
\nc{\ERGs}{ERGs}
\nc{\ibid}{\emph{ibid.}}
\nc{\viz}{viz}
\nc{\eg}{e.g.}
\nc{\ie}{i.e.}
\nc{\cf}{cf.}
\nc{\etc}{etc.}
\nc{\etcB}{etc}
\nc{\rhs}{r.h.s.}
\nc{\rhsB}{r.h.s}
\nc{\lhs}{l.h.s.}
\nc{\lhsB}{l.h.s}
\nc{\wrt}{with respect to}
\nc{\etal}{\emph{et al.}}
\nc{\aka}{a.k.a.}
\nc{\sic}{sic}
\nc{\QFT}{QFT}
\nc{\BS}{\Lambda_0}
\nc{\ES}{\Lambda}
\nc{\PV}{PV}
\nc{\ala}{\emph{\`a la}}
\nc{\role}{r\^{o}le}

\nc{\be}{\begin{equation}}
\nc{\ee}{\end{equation}}
\nc{\bea}{\begin{eqnarray}}
\nc{\eea}{\end{eqnarray}}
\nc{\beas}{\begin{eqnarray*}}
\nc{\eeas}{\end{eqnarray*}}

\nc{\bcf}{\begin{center}\begin{figure}}
\nc{\ecf}{\end{figure}\end{center}}

\nc{\bct}{\begin{center}\begin{table}}
\nc{\ect}{\end{table}\end{center}}

\nc{\ds}{\displaystyle}

\nc{\gap}{\hspace{.15em}}

\nc{\eq}[1]{(\ref{#1})}

\nc{\tr}{\mathrm{tr}\,}
\nc{\str}{\mathrm{str}\,}
\nc{\strI}{\mathrm{str} \!\! \int \!\! d^D \! x\,}
\nc{\ODInt}[1]{\int \!\! d#1}
\nc{\ODIntL}[3]{\int_{#2}^{#3} \!\! d#1}
\nc{\Int}[1]{\int \!\! d^D \! #1 \,}
\nc{\IntB}[1]{\int \!\! \frac{d^D \! #1}{(2\pi)^D} \,}
\nc{\DoubleInt}{\int\!\!\!\!\int}
\nc{\volume}[1]{d^D \! #1 \,}
\nc{\ve}[1]{d^D \! #1}
\nc{\AngVol}[1]{\not{\!\Omega_#1}}
\nc{\PowAngVol}[2]{\not{\! \! \Omega_#1^#2}}

\nc{\desl}{\partial \hspace{-.51em}/ }
\nc{\asl}{A \hspace{-.48em}/ }
\nc{\psl}{p \hspace{-.51em}/ }
\nc{\ksl}{k \hspace{-.51em}/ }
\nc{\Osl}{\not{\! \Omega}} 

\nc{\gam}{\gamma}
\nc{\Lam}{\Lambda}
\nc{\chibar}{\bar{\chi}}
\nc{\psibar}{\bar{\psi}}
\nc{\Phibar}{\bar{\Phi}}

\nc{\demu}{\partial_{\mu}}
\nc{\denu}{\partial_{\nu}}

\nc{\kernel}[1]{\!\cdot #1\!\cdot\!}
\nc{\covkernel}[1]{\{ #1 \}} 
\nc{\EP}[1]{\Delta^{#1}}
\nc{\EPB}[2]{\Delta^{#1}_{#2}}
\nc{\DEP}[1]{\dot{\Delta}^{#1}}
\nc{\DEPB}[2]{\dot{\Delta}^{#1}_{#2}}
\nc{\ci}{c^{-1}}
\nc{\hS}{\hat{S}}
\nc{\wrn}[2]{\gamma^{(#1)}_{#2}}

\nc{\comm}[2]{\left[ #1,#2 \right]}

\nc{\one}{\ensuremath{1\! \mathrm{l}}}
\nc{\der}[2]{\ensuremath{\frac{d #1}{d #2}}}
\nc{\pder}[2]{\ensuremath{\frac{\partial #1}{\partial #2}}}
\nc{\fder}[2]{\ensuremath{\frac{\delta #1}{\delta #2}}}
\nc{\order}[1]{\mathcal{O}\left( #1 \right)}
\nc{\Op}[1]{\mathcal{O}(p^{#1})}
\nc{\hf}{\frac{1}{2}}

\nc{\flow}{\ensuremath{-\Lambda \partial_\Lambda }}
\nc{\dec}[3][0]{\ensuremath{\left[ #2 \hspace{#1in} \right]^{#3}}}
\nc{\TLTP}[1]{S_{0 \mu \,\nu}^{\ AA}(#1)}
\nc{\OLDs}{\mathcal{D}_1}

\nc{\DAD}{\scriptstyle \odot}

\nc{\inte}{\!\int\!}
\nc{\dga}{\ensuremath{{\cal D}_{\Gamma}}}

\newlength{\LabLength}
\newlength{\ProcessRefLength}
\newlength{\ProcessLength}
\newlength{\CancelRefLength}
\newlength{\CancelLength}

\nc{\cd}[1]{\ensuremath{\begin{array}{c}\input{pstex/#1.pstex_t} \end{array}}} 

\nc{\sco}[3][0]{
	\begin{array}{c}
		#2 \\[#1ex]
		#3
\end{array}
}

\nc{\scd}[3][0]{
	\sco[#1]{\cd{#2}}{\cd{#3}}
}

\nc{\LD}[1]{
	\settowidth{\LabLength}{\scriptsize \textbf{\ref{#1}}}
	\addtolength{\LabLength}{0.8em}
	\begin{minipage}{\LabLength}
		\scriptsize
		\begin{D}\label{#1}\end{D}
	\end{minipage}
}

\nc{\LDLB}[1]{
	\settowidth{\LabLength}{\scriptsize \textbf{\{ \ref{#1}}}
	\addtolength{\LabLength}{0.8em}
	\begin{minipage}{\LabLength}
		\scriptsize
		\begin{DC}\label{#1}\end{DC}
\end{minipage}
}

\nc{\LDRB}[1]{
	\settowidth{\LabLength}{\scriptsize \textbf{\{ \ref{#1}}}
	\addtolength{\LabLength}{0.8em}
	\begin{minipage}{\LabLength}
		\scriptsize
		\begin{D}\label{#1}\textbf{\}} \end{D}
\end{minipage}
}

\nc{\PD}[2]{
	\settowidth{\LabLength}{\scriptsize\textbf{\ref{#1}}}
	\settowidth{\ProcessRefLength}{\scriptsize\ref{#2}}
	\addtolength{\LabLength}{\ProcessRefLength}
	\addtolength{\LabLength}{\ProcessLength}
	\addtolength{\LabLength}{0.8em}
	\begin{minipage}{\LabLength}
		\scriptsize
		\begin{D}\label{#1}$\rightarrow \ref{#2}$\end{D}
	\end{minipage}
}

\nc{\CD}[2]{
	\settowidth{\LabLength}{\scriptsize\textbf{\ref{#1}}}
	\settowidth{\CancelRefLength}{\scriptsize$\ref{#2}$}
	\addtolength{\LabLength}{\CancelRefLength}
	\addtolength{\LabLength}{\CancelLength}
	\begin{minipage}{\LabLength}
		\scriptsize
		\begin{DC}\label{#1} \ref{#2} \textbf{\}} \end{DC}
	\end{minipage}
}

\nc{\Discard}[1]{
	\settowidth{\LabLength}{\scriptsize\textbf{\ref{#1}}}
	\settowidth{\ProcessRefLength}{\scriptsize $\rightarrow 0$}
	\addtolength{\LabLength}{\ProcessRefLength}
	\addtolength{\LabLength}{\ProcessLength}
	\addtolength{\LabLength}{0.8em}
	\begin{minipage}{\LabLength}
	\scriptsize
		\begin{D}\label{#1}$\rightarrow 0$\end{D}
	\end{minipage}
}

\nc{\PO}[4][1]{
	\begin{array}{c}
		\PD{#3}{#4}
	\\[#1ex]
		#2
	\end{array}
}

\nc{\PDi}[4][1]{
	\PO[#1]{\cd{#2}}{#3}{#4}
}

\nc{\DO}[3][1]{
	\begin{array}{c}
		\Discard{#3}
	\\[#1ex]
		#2
	\end{array}
}

\nc{\Cancel}[2]{
\begin{cancel}
	Diagram~\ref{#1}  exactly cancels diagram~\ref{#2}.
	\label{cancel:#1}
\end{cancel}
}

\newcommand{\jhep}[3]{\emph{JHEP} #1 (#2) #3}
\newcommand{\NuclPhys}[4]{\emph{Nucl.\ Phys.\ }\textbf{#1 #2} (#3) #4}
\newcommand{\PhysRev}[4]{\emph{Phys.\ Rev.\ }\textbf{#1 #2} (#3) #4}
\newcommand{\IntJModPhys}[4]{\emph{Int.\ J.\ Mod.\ Phys.\ }\textbf{#1 #2} (#3) #4}
\newcommand{\PhysRep}[4]{\emph{Phys.\ Rep.\ }\textbf{#1 #2} (#3) #4}

\newcommand{\TheorMathPhys}[3]{\emph{Theor.\ Math.\ Phys.\ }\textbf{#1} (#2) #3}

\newcommand{\hepth}[1]{hep-th/#1}

\newcommand{\http}[1]{http://#1}
\newcommand{\arxiv}[1]{[#1]}

\title{Manifestly gauge invariant QED}

\author{
Stefano Arnone,${}^\P$ Tim R. Morris$\,\, {}^\dag$ and Oliver
J. Rosten${}^\dag$\\[0.5cm]
${}^\P$Dipartimento di Fisica, Universit\`a degli Studi di Roma ``La
Sapienza''\\
P.le Aldo Moro, 2 - 00185 Roma, Italy\\[0.2cm]
${}^\dag$School of Physics and Astronomy,  University of Southampton,\\
Highfield, Southampton SO17 1BJ, U.K.\\[0.5cm]
E-mails: {\tt Stefano.Arnone@roma1.infn.it, }\\
{\tt T.R.Morris@soton.ac.uk, O.J.Rosten@soton.ac.uk}
}
\date{}

\begin{document}

\maketitle

%%%\preprint{SHEP 02-22}  %works only in JHEP style
\begin{abstract}
We uncover a method of calculation that proceeds
at every step without fixing the gauge or specifying details of
the regularisation scheme. Results are obtained
%virtually without calculation,
by iterated use of integration by parts and gauge invariance
identities. Calculations can be performed almost entirely diagrammatically.
The method is formulated within the framework of an exact
renormalisation group for QED.
We demonstrate the technique with a calculation of the one-loop
beta function, achieving a manifestly universal result, and without
gauge fixing.
\end{abstract}

\vspace{-15truecm}
\begin{tabbing}
\hspace{9.5 truecm} \=CERN-PH-TH/2004-123\\
\>ROMA - 1399/04 \\
\>SHEP 04-21
\end{tabbing}

\newpage
\tableofcontents

\section{Introduction and Conclusions}      \label{sec:Intro}

The purpose of this paper is to illustrate a novel methodology
which  has been developed for Yang-Mills theory, in the simpler
context of (massless) \QED. In~\cite{YM-1-loop}, a manifestly
gauge invariant Exact Renormalisation Group (\ERG) was introduced,
suitable for computation in $SU(N)$ gauge theory, to one-loop,
without the need for gauge fixing. The formalism has now been
refined to facilitate computation to arbitrary loop
order~\cite{Quarks2004,Thesis}. Indeed, the (universal) $SU(N)$
Yang-Mills two-loop $\beta$-function has been successfully
calculated~\cite{Thesis}, representing the very first continuum
two-loop calculation to be done in Yang-Mills theory, without
fixing the gauge. This paper provides an introduction to these
ideas, whilst bypassing many of the technical subtleties. We will
see that it is thus straightforward to adapt these manifestly
gauge invariant methods to the case of perturbative computations
in QED.

To treat \QED\ within the framework of the manifestly gauge
invariant \ERG, we must
adopt a regularisation scheme which incorporates a gauge invariant,
real cutoff, $\Lambda$. Whereas this is a source of much
of the complication in the non-Abelian case~\cite{SU(N|N)},
it is straightforward for \QED. The pure gauge part of
the theory can be regulated by supplementing the kinetic
term with a cutoff function, $c$, which dies off
sufficiently fast at high energies. To regulate the fermionic
part of the action, we introduce an unphysical, commuting
spinor field, $\chi$, with a mass at the cutoff scale, which provides
Pauli-Villars regularisation.

When providing regularisation by Pauli-Villars fields,
one has to provide a prescription for adding together the
separately ultraviolet divergent pieces. Here this is easily
solved by the traditional method of momentum routing, \ie\ we
ensure that the loop momentum assignments for every $\chi$ loop are
the same as the assignments in the corresponding physical fermion
loop~\cite{SU(N|N)}.
In fact, as we will see, such a routing is embedded in the techniques described in
this paper since the corresponding diagrams are represented as one
diagram, with internal fields summed over.

However, to reflect the more involved case of a flow equation for
$SU(N)$ Yang-Mills theory where the $SU(N|N)$ regularisation
requires a preregularisation,
we use dimensional regularisation as a
preregularisation.\footnote{There are encouraging
indications~\cite{Thesis} that an entirely diagrammatic
prescription can instead be adopted for $SU(N)$ Yang-Mills theory,
which would make sense not only
in general $D$ but also in $D$
strictly equal to four.}  Working in general
dimension, $D$,  will prove particularly convenient as
this allows an efficient means of
extracting the numerical value of $\beta$-function coefficients
(see section~\ref{sec:Numerics}).
The regularisation scheme
is fully described in appendix~\ref{app:Reg}.

Having cast \QED\ in a form suitable for the \ERG\ in section~\ref{sec:adapt},
in section~\ref{sec:FlowEquation}
we formulate the flow equation, which describes how the
effective action, $S \equiv S_\Lambda$, changes with $\Lambda$. Using
the insights of~\cite{TRM+JL-0} into general \ERGs\ and our
experiences in both the non-Abelian case~\cite{YM-1-loop}
and scalar case~\cite{scalar2loop}, an appropriate equation
can simply be written down. Indeed, there are an infinite
number of flow equations we can use, from each of which we
can extract equivalent physics. The key point is that
all of these equations must satisfy the following properties.

First, the flow equation must be gauge invariant. That we are able
to formulate gauge invariant \ERGs\ turns out to have a wonderful
benefit: calculations can be performed without gauge
fixing~\cite{GI-ERG-I,YM-1-loop,Thesis}. In \QED\ (and non-Abelian
gauge theory), the exact preservation of gauge invariance means
that the connection does not renormalise. By scaling the coupling
out of the connection, the gauge field does not renormalise
(though we note that the fermion and unphysical regulator fields
do suffer wavefunction renormalisation).

Secondly, all ingredients of the flow equation must be infinitely
differentiable in momenta. This property, referred to
as quasilocality~\cite{TRM-MassiveScalar}, guarantees that
each RG step $\Lambda \mapsto \Lambda - \delta \Lambda$
does not generate IR singularities~\cite{Wilson}; equivalently
that Kadanoff blocking takes places only over a localised patch.

Thirdly, the flow is self-similar~\cite{SelfSimilarFlow}, which
demands that the effective action depend only on the scale
$\Lambda$, through its couplings (this can be straightforwardly
extended to massive theories~\cite{TRM-MassiveScalar}). By
definition, such actions lie on a Renormalised Trajectory
guaranteeing, amongst other things, the existence of a continuum
limit~\cite{TRM-elements}.

Lastly, the partition function must be left invariant under the flow:
$e^{-S}$ should transform into a total derivative.
This ensures that, starting with some quantum field theory at a high scale,
our effective action is guaranteed to still be describing the same physics
at a very much lower scale (implicitly assuming the second point, above).

In turn, these features imply that our flow equation possesses
other properties. Indeed, it follows that our flow equation
does correspond to integrating out~\cite{GI-ERG-I,Quarks2004}.
Furthermore, locality and that we are really dealing with \QED\ at all
scales---two properties which are
generally not manifest in the Wilsonian effective action at some finite value of
$\Lambda$---are guaranteed (at least in perturbation theory)~\cite{YM-1-loop}.

%Due to invariance of the partition function under the flow
%(and the fact that the flow is free of IR divergences), we know that,
%if we are dealing with a local action at some scale, then we must
%have been dealing with a local action at all higher scales and are
%guaranteed to be dealing with a local action at all lower scales.
%Now, we know that the flow lies on a
%renormalised trajectory and that this trajectory is controlled by
%the free-field (\aka\ Gaussian) fixed point. To obtain a continuum
%limit, the flow is tuned such that it passes arbitrarily close
%to the free-field fixed point. At this point, the action---being
%that of a non-interacting field theory---is local. Hence, locality
%is guaranteed at all other points along the flow.

%Similarly, we can argue that we are always dealing with \QED,
%even though we generically expect the effective action to be
%some complicated functional of not just the physical fields but also of
%the unphysical regulator: we know that, at sufficiently high
%energies the regulator field decouples; since we are dealing
%with \QED\ here, we are guaranteed to be dealing with \QED\ everywhere
%else along the flow.

%It is worth commenting on the end of the renormalised
%trajectory, where $\ES \rightarrow \infty$.
%The action here is just the bare action, whose form is determined by
%the flow equation, but whose precise details amount to choices we are free to make.
%As is generally our philosophy, we leave these choices unmade.

When we explicitly construct the flow equation (see
equations~\eq{eq:FlowSchematic}--\eq{eq:Flow-a1}), we will see
that it depends not just on the Wilsonian effective action but
also on a second functional $\hat{S}$, the `seed
action'~\cite{YM-1-loop,scalar2loop}. This object controls the
flow and represents the continuum version of the choice of
blocking transformations in the application of Wilsonian RG
techniques to latticised problems~\cite{Wilson,scalar1loop}. Like
the Wilsonian effective action, the seed action must be gauge
invariant and infinitely differentiable. However, whereas we solve
the flow equations for the former, the latter acts as an input.

Now, so long as our choice of $\hat{S}$ is consistent with our approach,
we are free to choose it to be whatever we like. However, we expect
the result of a computation of some universal quantity to be independent
of this choice and, indeed, our choice of cutoff function, $c$.

That a calculation must yield a result independent of the precise details
of particular ingredients
leads to a highly constrained calculational procedure. Generally speaking,
the only way to remove dependence on an instance of an unspecified
component
of $\hat{S}$ is for there to be a second instance, of opposite sign. Indeed,
this is so constraining that calculations can be performed almost entirely
diagrammatically.

Whilst the freedom to leave the seed action largely unspecified has guided
us to an efficient calculational procedure we have, in some sense, complicated
the issue by not using the simplest seed action suitable for our purposes.
The most obvious resolution to this is to simply regard the freedom
of the seed action as scaffolding: it has guided us to an efficient calculational
procedure, but now we can dispense with it, keeping the procedure but choosing
the simplest form for the seed action, consistent with our approach.

However, there is a more sophisticated way of looking at things.
Certainly, for the calculation of $\beta$-function
coefficients---\emph{to any order in perturbation theory}---it is
possible to show that explicit dependence
on $\hat{S}$ is guaranteed to cancel out~\cite{Thesis,GeneralMethods}.
Thus, at least up until this stage of a
calculation, where any remaining seed action dependence is now implicit,
it makes no difference how complicated a seed action has been chosen; we
can simply bypass the entire procedure of cancelling explicit
instances of seed action components.

A second benefit of a general seed action is that, by making
certain judicious choices, we can simplify the calculation
procedure employed within the \ERG. Specifically, we choose to
set the classical, two-point Wilsonian effective action vertices
equal to the corresponding seed action vertices. A direct consequence
of this is that the classical two-point vertices are now related,
in a particularly simple way, to the integrated \ERG\ kernel. This leads
to the so-called `effective propagator relationship', which is
central to our entire diagrammatic approach.

The diagrammatics are introduced in section~\ref{sec:Diagrammatics}
and then specialised to the weak coupling regime in sections~\ref{sec:WeakCoupling}
and~\ref{sec:EffProp}. This prepares us for the final section, \ref{sec:beta1}, in which
we illustrate how to use the formalism to perform actual calculations, by computing
the \QED\ one-loop $\beta$-function, without fixing the gauge or
specifying the non-universal details of either $\hS$ or $c$.

\section{Adapting \QED\ for the ERG}        \label{sec:adapt}

We denote the gauge field by $A_\mu$, the physical fermion field by $\psi$
and the unphysical regulating field by $\chi$. The
covariant derivative is
\be
    \nabla_\mu = \partial_\mu - i A_\mu,
\label{eq:CovariantDerivative}
\ee
where it is apparent that we have scaled the renormalised coupling, $e$, out of this definition.
The renormalised coupling, $e$, and the
canonical normalisation of the spinor fields are defined through the renormalisation condition
\be
    S = \frac{1}{e^2} \Int{x}
        \left(
            \frac{1}{4} F_{\mu \nu}^2 +\psibar i \desl \psi + \chibar i \desl \chi + \cdots,
        \right)
\label{eq:RenormCondition}
\ee
where $F_{\mu \nu} = i \comm{\nabla_\mu}{\nabla_\nu} = \demu A_\nu - \denu A_\mu$
is the field strength and the ellipsis
denotes higher dimension operators (and the ignored vacuum energy).
The mass of the physical fermion is set to zero implicitly by ensuring that
there is no other
physical scale apart from $\Lambda$, which itself tends to zero as all momentum
modes are integrated out.

The effective
action can be written
\be
    S = \sum_{n=2}^\infty \frac{1}{s_n} \Int{x_1} \!\! \cdots \volume{x_n}
        S^{X_1 \cdots X_n}_{\,a_1 \, \cdots \, a_n}(x_1, \ldots, x_n) X_1^{a_1}(x_1) \cdots X_n^{a_n} (x_n),
\label{eq:EffectiveAction}
\ee
where $X$ represents any of $A_\mu$, $\psi$ or $\chi$; the indices $a_i$ being
Lorentz indices or spinor indices, as appropriate. Each vertex must possess a minimum of two fields
and those vertices possessing instances of $\psi(\chi)$ without a matching
 $\psibar(\chibar)$ are given a value of zero. For each list, $X_1 \cdots X_n$,
only one permutation of the fields appears in the sum;
moreover, we will take the anticommuting $\psi$, $\psibar$s to be
\emph{canonically ordered}, by which we mean they always occur
in the order $\psibar \psibar \cdots \psi \psi$. The symmetry factor of each vertex
is $s_n = \prod_f n_f!$, where $n_f$ is the number of fields of type $f$
and $f = A, \ \psi, \ \psibar, \ \chi, \ \chibar$.

We write the momentum space vertices as
\begin{eqnarray*}
    \lefteqn{
        S^{X_1 \cdots X_n}_{\,a_1 \, \ldots \, a_n}(p_1, \ldots, p_n) \left(2\pi \right)^D \delta \left(\sum_{i=1}^n p_i \right)
    }
\\
    & &
    =
    \Int{x_1} \!\! \cdots \volume{x_n} e^{-i \sum_i x_i \cdot p_i} S^{X_1 \cdots X_n}_{\,a_1 \, \ldots \, a_n}(x_1, \ldots, x_n),
\end{eqnarray*}
where all momenta are taken to point into the vertex. We will employ the shorthand
\be
    S^{X_1 X_2}_{\, a_1 \, a_2}(p) \equiv S^{X_1 X_2}_{\, a_1 \, a_2}(p,-p).
\label{eq:TwoArgs}
\ee

The effective action~\eq{eq:EffectiveAction} is
invariant under the gauge transformation
\bea
    \delta \psi     &=&  i \alpha(x) \psi, \label{eq:GT-psi}
\\
    \delta \psibar  &=& -i \psibar \alpha(x),\label{eq:GT-psibar}
\\
    \delta \chi     &=&  i \alpha(x) \chi, \label{eq:GT-chi}
\\
    \delta \chibar  &=& -i \chibar \alpha(x),\label{eq:GT-chibar}
\\
    \delta A_\mu    &=& \demu \alpha(x). \label{eq:GT-A}
\eea

Notice that by scaling the coupling constant out of~\eq{eq:CovariantDerivative},
we ensure that $A_\mu$ does not suffer from any
wavefunction renormalisation.
If $A_\mu$ were to acquire a wavefunction
renormalisation then, on replacing $A_\mu$ by $Z^{1/2} A_\mu$,
it is clear that,
in order not to violate
\eq{eq:GT-A}, we must re-parametrise $\alpha$ as well. However,
re-parameterising $\alpha$ is inconsistent with~\eq{eq:GT-psi}--\eq{eq:GT-chibar}
even taking into account the
wavefunction renormalisation for the physical
fermion and unphysical regulator field.

Invariance of the effective action, \eq{eq:EffectiveAction},
under the transformations~\eq{eq:GT-psi}--\eq{eq:GT-A}
implies the following Ward identities.
First, pure-$A$ vertices are transverse on all legs:
\be
        p_i^{\mu_i}\, S^{A \; \cdots A \; \cdots A}_{\mu_1 \cdots \mu_i \cdots \mu_m} (p_1,\ldots,p_i,\ldots,p_m)
        = 0, \qquad \forall p_i^{\mu_i}.
\label{eq:WI-A}
\ee

Secondly, vertices containing $\psi$s and / or $\chi$s and
at least one gauge field are related to vertices with one
less gauge field. It is useful to define the field $\Phi$ to be either
$\psi$ or $\chi$; in the former (latter) case, $\Phibar$
should be identified with $\psibar$ ($\chibar$). The Ward
identity is:
\bea
    \lefteqn{k^\mu S^{A A\; \cdots A \ \Phibar_1 \cdots\Phibar_m \Phi_1 \cdots\Phi_m}_{\mu \, \mu_1 \cdots \mu_n a_1\, \cdots a_m \; b_1 \, \cdots \, b_m}
        (k,p_1,\ldots,p_n,q_1,\ldots,q_{m},r_1,\ldots, r_m) = }\nonumber
\\[1ex]
    & &\sum_i \bigg\{  S^{A\; \cdots A \ \Phibar_1 \cdots\Phibar_m \Phi_1 \cdots \Phi_i \cdots \Phi_m}_{
        \mu_1 \cdots \mu_n a_1\, \cdots a_m \; b_1 \, \cdots \, b_i \, \cdots\,  b_m}
        (k,p_1,\ldots,p_n,q_1,\ldots q_m,r_1, \ldots, r_i +k, \ldots, r_m)\nonumber
\\[1ex]
    & & -  S^{A\; \cdots A \ \Phibar_1 \cdots \Phibar_i \cdots \Phibar_m \Phi_1 \cdots \Phi_m}_{
        \mu_1 \cdots \mu_n a_1\, \cdots \, a_i \, \cdots \, a_m \; b_1 \, \cdots \, b_m}
         (k,p_1,\ldots,p_n,q_1,\ldots,q_i + k,\ldots q_{m},r_1\ldots r_m) \bigg\}.
\label{eq:WI-F}
\eea
The above generalises the well-known relation, $p^\mu S^{A \psibar \psi}_\mu
(p,q,r) =S^{\psibar \psi}(q) - S^{\psibar \psi}(r)$, where the
shorthand~\eq{eq:TwoArgs} has been used.

In addition to being invariant under gauge transformations, the action is
charge conjugation invariant. In particular, this implies that any pure-$A$
vertices carrying an odd number of fields vanish.

The seed action, $\hS$, has a field expansion as in~\eq{eq:EffectiveAction}
and obeys the same symmetries as the Wilsonian effective action.

\section{An \ERG\ for \QED}     \label{sec:ERGforQED}

\subsection{The Flow Equation}  \label{sec:FlowEquation}

Our strategy is to write down a manifestly gauge invariant flow
equation that incorporates the regularisation scheme outlined
above. Following our previous works
\cite{YM-1-loop,Quarks2004,Thesis,scalar2loop,GI-ERG-I,scalar1loop,GI-ERG-II},
we simply set (suppressing spinor indices)
\be
    \dot{S} + \gamma^{(\psi)} \left( \psi \fder{S}{\psi} + \psibar \fder{S}{\psibar} \right)
            + \gamma^{(\chi)} \left( \chi \fder{S}{\chi} + \chibar \fder{S}{\chibar} \right)
            = a_0[S,\Sigma_e] - a_1[\Sigma_e],
\label{eq:FlowSchematic}
\ee
where $\dot{S} \equiv \flow S$ (in general dots above quantities will
signify $\flow $), the $\gam$s take into account the contribution of the
anomalous dimension of the spinor fields and $\Sigma_e \equiv e^2 S - 2
\hS$.  The functionals $a_{0,1}$ are given by:
\bea
    a_0[S, \Sigma_e]    & = & \frac{1}{2} \fder{S}{A_\mu} \kernel{\DEP{AA}} \fder{\Sigma_e}{A_\mu}
                            + \frac{1}{2}
                            \left(
                                \fder{\Sigma_e}{\chi} \covkernel{\DEP{\chibar \chi}} \fder{S}{\chibar} +
                                \fder{S}{\chi} \covkernel{\DEP{\chibar \chi}} \fder{\Sigma_e}{\chibar}
                            \right) \nonumber
\\[1ex]
                        &   &
                            \qquad - \frac{1}{2}
                            \left(
                                \fder{\Sigma_e}{\psi} \covkernel{\DEP{\psibar \psi}} \fder{S}{\psibar} +
                                \fder{S}{\psi} \covkernel{\DEP{\psibar \psi}} \fder{\Sigma_e}{\psibar}
                            \right) \label{eq:Flow-a0}
\\[1ex]
    a_1[\Sigma_e]       & = & \frac{1}{2} \fder{}{A_\mu} \kernel{\DEP{AA}} \fder{\Sigma_e}{A_\mu}
                            +
%                           \left(
%                               \fder{}{\chibar} \covkernel{\DEP{\chibar \chi}} \fder{\Sigma_e}{\chi} +
                                \fder{}{\chi} \covkernel{\DEP{\chibar \chi}} \fder{\Sigma_e}{\chibar}
%                           \right) \nonumber
                            -
%                           \left(
%                               \fder{}{\psibar} \covkernel{\DEP{\psibar \psi}} \fder{\Sigma_e}{\psi} -
                                \fder{}{\psi} \covkernel{\DEP{\psibar \psi}} \fder{\Sigma_e}{\psibar}
%                           \right)
                            \label{eq:Flow-a1}.
\eea

To define these expressions properly, we must define our notation. Let us start
by looking at terms involving functional derivatives \wrt\ $A$.
Under the gauge transformations~\eq{eq:GT-psi}--\eq{eq:GT-A},
both $\delta S / \delta A_\mu$ and $\delta \Sigma_e / \delta A_\mu$ are invariant.
Thus, construction of  a gauge invariant term in the flow equation involving these
objects is trivial. We write the $A$-sector term in compact form
by utilising the following shorthand:
for any two functions $f(x)$ and $g(y)$ and a
momentum space kernel $W(p^2/\Lambda^2)$,
\be
    f \kernel{W} g = \DoubleInt \volume{x} \volume{y} f(x)\, W_{x y}\,g(y),
\label{eq:fWg}
\ee
with $W_{x y}$ being $\IntB{p} W(p^2/\Lambda^2) e^{i p \cdot (x-y)}$.

In the $\psi$ and $\chi$-sectors, however, things are more complicated
since $\delta S / \delta \chi$ and so on are not invariant under
gauge transformations. To construct gauge invariant terms for
the flow equation, we must covariantise~\eq{eq:fWg}.
Thus, for the spinors $u_a(x)$ and $v_b(y)$ and kernel $W_{ab}$, the gauge invariant generalisation
of~\eq{eq:fWg} is~\cite{GI-ERG-I}
\bea
    \lefteqn{u \covkernel{W} v = \sum_{n=0}^{\infty} \Int{x}\volume{y}\volume{x_1}\!\!\cdots\volume{x_n} } \nonumber
\\
    &&
    u_a(x) W_{ab \; \mu_1\cdots\mu_n}(x_1,\ldots,x_n;x,y) A_{\mu_1}(x_1) \cdots A_{\mu_n}(x_n) v_b(y),
\label{eq:Wine}
\eea
where the $n$-point vertex $W_{ab \; \mu_1\cdots\mu_n}(x_1,\ldots,x_n;x,y)$ is
a vertex of the
covariantised kernel, $\covkernel{W}$.
If $n=0$, we recover the original kernel: $W_{ab}(;x,y) \equiv W_{ab}(x,y)$.

In the flow equation, where $u$ and $v$ both involve a functional derivative,
we use the field with respect to which this derivative is taken to label
the kernels. The vertices of the \ERG\ covariantised kernels
$\covkernel{\DEP{XY}}$ obey Ward identities similar to~\eq{eq:WI-F},
though we will not require them in this paper.

The final point to make about the flow equation is that we can choose the
kernel $\DEP{\psibar \psi}$ to be zero, which clearly simplifies the
flow equation.  Nonetheless, even given this choice,
it is useful to retain these terms and to process them
using the flow equation. As we will see when we describe the diagrammatics,
this does not actually lead to any extra work since all sectors can be treated in parallel.
 Moreover,
allowing the $\psi$-sector to shadow the $\chi$-sector will enable us to
extract actual numerical answers from the flow equation in a convenient manner
by making the UV regularisation of the results manifest.

\subsection{Diagrammatics}      \label{sec:Diagrammatics}

In this section, we will give a diagrammatic form for
the flow equation. First, though, we must introduce the
diagrammatics for the actions and the kernels. Referring back
to~\eq{eq:EffectiveAction}, consider the vertex with $n$ gauge fields.
We represent the \emph{vertex coefficient function}---\ie\ we do not
include either the actual fields or the symmetry factor---as shown
in figure~\ref{fig:Diags:Vertices}. We henceforth refer to all lines
emanating from a vertex as decorations.
\bcf[h]
    \[
        S^{A \; \cdots A}_{\mu_1 \cdots \mu_n}(p_1, \ldots, p_n)  =\cd{Vertex-S-A1--An}
    \]
\caption{Example of the diagrammatic representation of a vertex coefficient
function.}
\label{fig:Diags:Vertices}
\ecf

The argument of the vertex can be any action functional and
so the vertices of $\hS$ (or, indeed, $\Sigma_e$) have
a similar diagrammatic interpretation.
Inclusion of $\chi$s
 is straightforward.
However, the order in which anticommuting fermionic
decorations are read off is important:
we always read off such that
the algebraic form of the vertex is canonically ordered.
Since we will
never need to deal with external $\psi$s or $\chi$s in this paper,
we need not introduce a specific diagrammatic realisation. Rather,
we will use solid lines to indicate dummy fields; these will
appear as internal lines in diagrams, whence the possible
fields they represent are summed over.

The diagrammatics for the vertices of the kernels is shown in
figure~\ref{fig:WineDiagrammatics}; this time, we have not explicitly
indicated the momentum flow but, once again, all momenta are taken to flow
into the diagram. From~\eq{eq:Wine} we note that only $A$s decorate
the kernel.
\bcf[h]
    \[
        \dot{\Delta}^{X\;Y}_{a_1 a_2 \; \mu_1\cdots\mu_n}(p_1,\ldots,p_n;r,s) = \cd{WineVertex}
    \]
\caption{Diagrammatics for the vertices of the kernels}
\label{fig:WineDiagrammatics}
\ecf

We now refine the diagrammatics. Consider the vertex coefficient
function corresponding to the vertex decorated by some set of
fields $\{f\}$. Rather than explicitly performing these decorations,
it is useful to leave them implicit, as shown in figure~\ref{fig:Diags:ImplicitDecoration}.
\bcf[h]
    \[
        \dec{
            \cd{Vertex-S}
        }{\{f\}}
    \]
\caption{A Wilsonian effective action vertex implicitly decorated by the set of fields $\{f\}$.}
\label{fig:Diags:ImplicitDecoration}
\ecf

Using the notation of figure~\ref{fig:Diags:ImplicitDecoration},
the flow equation can now be
cast in a particularly simple diagrammatic form,
shown in figure~\ref{fig:Diags:FE}. The terms on the \rhs\ are formed
by $a_0$ and $a_1$, respectively.
\bcf[h]
    \[
        \dec{
            \cd{Vertex-S-LdL} +2\left(n_\psi \gamma^{(\psi)} + n_\chi \gamma^{(\chi)} \right) \cd{Vertex-S}
        }{\{f\}}
        = \frac{1}{2}
        \dec{
            \cd{Dumbbell-S-Sig_e} - \cd{Padlock-Sig_e}
        }{\{f\}}
    \]
\caption{A diagrammatic representation of the flow equation.}
\label{fig:Diags:FE}
\ecf

The internal lines correspond to vertices of the kernels,
where we sum over the fields at both ends.
Decorating the diagrams of figure~\ref{fig:Diags:FE}
is straightforward: we
distribute the decorative fields $\{f\}$ in all
independent ways over all structures. Note that, for particular
choices of $\{f\}$, certain explicitly decorated structures will not exist;
equivalently, we take the Feynman rule for such structures to be
zero. For
example, we know that the vertices of the kernel cannot
be decorated by spinor fields.

Finally, we note that there are some additional rules for the term
formed by $a_1$. If the internal line is in the $\chi$-sector,
we pick up a factor of two whereas, if it is in the $\psi$-sector,
we pick up a factor of minus two.

By applying the flow equation iteratively (as we will do shortly) we
can generate a diagram formed by the action of an arbitrary number
of $a_1$s and $a_0$s. The Feynman rule for working out the sign of
such arbitrarily complicated diagrams associated with internal $\psi$s
is simple and as expected: we pick up a sign for every loop.
Note that loops can only ever be formed by $a_1$.

As an example, we compute the flow of a vertex decorated by two $A$s
(we will explicitly perform the decorations, rather than leaving them
implicit), as shown in figure~\ref{fig:FlowEx}.
\bcf[h]
    \[
    \dec{
        \cd{Vertex-AA}
    }{\bullet}
    =
    \frac{1}{2}
    \left[
        2 \cd{Dumbbell-S-A-Sig_e-A} - \cd{Padlock-Sig_e-AA} - 2\cd{Padlock-Sig_e-A-W-A} - \cd{Padlock-Sig_e-W-AA}
    \right]
    \]
\caption{Example of using the diagrammatic flow equation.}
\label{fig:FlowEx}
\ecf

Let us start by looking at the first term on the \rhsB. The
internal line whose flavour, we recall, is summed over, must
be in the $A$-sector. This diagram, like the third diagram
on the \rhs, comes with a relative factor of
two, arising from combining the diagrams
drawn as shown with those for which the $A^1_\mu$
and $A^1_\nu$ are swapped around (the two
cases must be the same, by Bose symmetry). Note that
no other diagrams can be formed by $a_0$, since these
would necessarily possess a one-point vertex, which does
not exist.

Let us now focus on the second term. The kernel
can be in any sector. If we suppose that
it is in the $\psi$-sector, then
our previous
rules determine the vertex to be $\Sigma_e^{\ \psibar \psi AA}$
and demand that we pick up a factor of $-2$. The third and fourth
terms only exist when the internal line is in the $\chi$ or $\psi$-sectors.

We now introduce an additional piece of notation
which allows us to perform an intermediate step
between going from unrealised decorations
parameterised by $\{f\}$ to explicitly decorated diagrams.

\begin{center}
\begin{figure}[h]
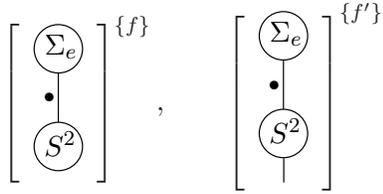

    \[
        \dec{
        \cd{Dumbbell-S2-Sigma_e}
    }{\{f\}}, \hspace{2em}
    \dec{
        \cd{Dumbbell-S2-Explicit-Sigma_e}
    }{\{f'\}}
    \]
\caption{Forcing a vertex on an implicitly decorated diagram
to have precisely two decorations.}
\label{fig:SpecificNumDecs}
\end{figure}
\end{center}

Suppose that we have a diagram for which we want to focus
on the components possessing a two-point vertex. So long
as there is still at least one field sitting as
an implicit decoration, we must specify that our vertex
has precisely two decorations. We use a superscript on
the vertex argument to denote this. Two
examples are shown in figure~\ref{fig:SpecificNumDecs}.

In the first diagram, we see that we must use one
and only one element of the set
$\{f\}$ to decorate the bottom vertex. In
the second diagram, the elements of $\{f'\}$ must be
spread around the kernel and the top vertex.

\section{The Weak Coupling Expansion}   \label{sec:WeakCoupling}

\subsection{The Weak Coupling Flow Equation}

In this section, we examine the form the flow equations take in
the perturbative limit. Following~\cite{GI-ERG-I,YM-1-loop}, the
action has the weak coupling expansion
\be
    S = \sum_{i=0}^\infty \left( e^2 \right)^{i-1} S_i = \frac{1}{e^2}S_0 + S_1 + \cdots,
\label{eq:WeakCouplingExpansion-Action}
\ee
where $S_0$ is the classical effective action and the $S_{i>0}$
the $i$th-loop corrections. The seed action has a similar expansion:
\begin{equation}
    \hS = \sum_{i=0}^\infty  e^{2i}\hat{S}_i.
\label{eq:WeakCouplingExpansion-SeedAction}
\end{equation}
Note that these definitions are consistent with $\Sigma_e = e^2 S -2\hat{S}$;
identifying powers of $e$ in the flow equation, it is clear that
$S_i$ and $\hat{S}_i$ will always appear together.
With this in mind, we now define
\be
    \Sigma_i = S_i - 2\hat{S}_i.
\label{eq:WeakCouplingSigma}
\ee
The $\beta$-function takes the usual form:
\be
    \beta \equiv \flow e = \sum_{i=1}^\infty  e^{2i+1} \beta_i.
\label{eq:WeakCouplingExpansion-beta}
\ee
and $\wrn{\psi}{}$, $\wrn{\chi}{}$ have the expansions
\be
    \wrn{\psi}{} = \sum_{i=1}^{\infty} e^{2i} \wrn{\psi}{i}, \qquad \wrn{\chi}{} = \sum_{i=1}^{\infty} e^{2i} \wrn{\chi}{i}.
\label{eq:WeakCouplingExpansion-gamma}
\ee

We now substitute expansions~\eq{eq:WeakCouplingExpansion-Action},
\eq{eq:WeakCouplingExpansion-SeedAction}, \eq{eq:WeakCouplingExpansion-beta}
and~\eq{eq:WeakCouplingExpansion-gamma}
into equation~\eq{eq:FlowSchematic} and, using equation~\eq{eq:WeakCouplingSigma},
obtain the weak coupling expansion of the flow equation:
\bea
&
\ds
    \dot{S}_n   = \sum_{r=1}^n
                    \left[
                        2 (n-r-1)\beta_r - \wrn{\psi}{r} \left( \psi \fder{}{\psi} + \psibar \fder{}{\psibar} \right) -
                                        \wrn{\chi}{r} \left( \chi \fder{}{\chi} + \chibar \fder{}{\chibar}  \right)
                    \right]S_{n-r} & \nonumber
\\
&
\ds
                     + \sum_{r=0}^n a_0 \left[ S_{n-r}, \Sigma_r \right] - a_1 \left[\Sigma_{n-1} \right].
&
\label{eq:WeakCouplingExpansion-Flow}
\eea

We note that the classical term can be brought
into a more symmetrical form. This follows from the invariance of
$a_0[S_{n-r}, \Sigma_r] + a_0[S_{r}, \Sigma_{n-r}]$
under $r \rightarrow n-r$. We exploit this by recasting
the classical term as follows:
\begin{eqnarray}
    a_0[\bar{S}_{n-r}, \bar{S}_r]   & \equiv    & a_0[S_{n-r}, S_r] - a_0[S_{n-r}, \hat{S}_r] - a_0[\hat{S}_{n-r}, S_r].
\label{eq:NFE:BarNotation}
\\
                                    & =         &   \left\{
                                                        \begin{array}{cc}
                                                            \frac{1}{2} \left(a_0[S_{n-r}, \Sigma_r] + a_0[S_{r}, \Sigma_{n-r}]\right)  & n-r \neq r
                                                        \\
                                                            a_0[S_r, \Sigma_r]                                                      & n-r =r.
                                                        \end{array}
                                                    \right.
\nonumber
\end{eqnarray}

Hence, we can rewrite the flow equation in the following form:
\bea
&
\ds
    \dot{S}_n = \sum_{r=1}^n    \left[
                                    2(n-r-1) \beta_r  - \wrn{\psi}{r} \left( \psi \fder{}{\psi} + \psibar \fder{}{\psibar} \right)
                                    - \wrn{\chi}{r} \left( \chi \fder{}{\chi} + \chibar \fder{}{\chibar}    \right)
                                \right]S_{n-r}
& \nonumber
\\
&
\ds
        + \sum_{r=0}^n a_0 \left[\bar{S}_{n-r}, \bar{S}_r \right] -  a_1 \left[\Sigma_{n-1} \right].
&
\label{eq:WeakCouplingExpansion-Flow-B}
\eea

When expressing this equation diagrammatically, we will abbreviate
the argument of Wilsonian
effective action vertices to just the loop order \ie\ we replace
$S_m$ by just $m$. Similarly, we represent seed action vertices by,
\eg, $\hat{m}$.

\subsection{The Effective Propagator Relation}  \label{sec:EffProp}

Central to our diagrammatic approach is
the relationship between the two-point, tree
level vertices and the integrated \ERG\ kernels---the `effective
propagator relation'.
To investigate this, we use the flow equations. First,
we specialise equation~\eq{eq:WeakCouplingExpansion-Flow}
or~\eq{eq:WeakCouplingExpansion-Flow-B} to $n=0$:
\be
    \dot{S}_0 = a_0[S_0, \Sigma_0].
\label{eq:TreeLevelFlow}
\ee
Next we further specialise to consider the flow of all
two-point vertices,
as shown in figure~\ref{fig:NFE:TLTP-flow}.
\bcf[h]
    \be
        \cd{Vertex-TLTP-LdL} = \cd{Dumbbell-S_0-Sigma_0}.
    \label{eq:TLTP-flow}
    \ee
\caption{Flow of all possible two-point, tree level vertices.}
\label{fig:NFE:TLTP-flow}
\ecf

We can now solve equation~\eq{eq:TLTP-flow}, giving an
expression for the Wilsonian effective action, two-point,
tree level vertices in terms of the seed action two-point,
tree level vertices and the zero-point kernels.

Our strategy is to follow~\cite{YM-1-loop,scalar2loop} and utilise
the freedom inherent in $\hat{S}$ by choosing the two-point, tree
level components of $\hat{S}$ to be equal to the corresponding
components of $S$ \ie\
\be
    \hat{S}^{\ \,f\, f}_{0 RS}(k) = S^{\ \,f\, f}_{0 RS}(k),
\label{eq:TLTPs=}
\ee
where the indices $R$ and $S$ are either Lorentz indices
or spinor indices, as appropriate.
We emphasise that~\eq{eq:TLTPs=}
is simply a choice we make, consistent with the flow
equations, that turns out to be helpful.

Given this choice, we can now uncover the effective
propagator relation. Using~\eq{eq:TLTPs=} and~\eq{eq:WeakCouplingSigma}, we note
that~\eq{eq:TLTP-flow} simplifies, since $\Sigma_{0 RS}^{\ \,f\, f} = - S_{0 RS}^{\ \,f\, f}$.
In the $\chi$-sector, equation~\eq{eq:TLTP-flow} becomes
\be
    \dot{S}^{\ \, \chibar \chi}_{0 \, a\, b}(p) = - S^{\ \, \chibar \chi}_{0 \, a \, c}(p) \DEPB{\chibar \chi}{c \, d}(p) S^{\ \, \chibar \chi}_{0 \, d\, b}(p),
\label{eq:TLTP-chi}
\ee

Integrating up, we
set the integration constant to zero to ensure
that the integrated kernel for the massive field $\chi$ is well
behaved as
$p \rightarrow 0$. This yields:
\be
    S^{\ \, \chibar \chi}_{0 \, a \, c}(p) \EPB{\chibar \chi}{c \, b}(p)  = \delta_{ab}.
\label{eq:EP-chi}
\ee
In other words, the integrated $\chi$ kernel is the inverse of
the two-point, tree level vertex, in the $\chi$-sector. In recognition of
this relationship, we call $\EP{\chibar \chi}(p)$ an effective
propagator. Indeed, in this sector of the calculation the effective
propagator is
essentially a propagator in the usual sense. However, that this is
the case is due to a succession of choices we made purely
for convenience. Moreover, in the $A$-sector we cannot even
define a propagator in the usual sense, since we have not fixed the gauge.
Consequently, we use the terminology `effective propagator' to highlight
the similarity of these object to normal propagators, mindful that
this relationship should not be taken for granted.

In the $A$-sector, matters are slightly more complex.
Equation~\eq{eq:TLTP-flow} becomes
\be
    \dot{S}^{\ AA}_{0\mu\, \nu} (p) = -  S^{\ AA}_{0\mu\, \alpha} (p) \DEP{AA}(p) S^{\ AA}_{0\alpha\, \nu} (p).
\label{eq:TLTP-A}
\ee
Now, from
the Ward identity~\eq{eq:WI-A},
the two-point, tree level vertex must be transverse. Thus, by
dimensions, it must take the form
\be
    S^{\ AA}_{0 \mu\, \nu} (p) = \frac{\Box_{\mu \nu}(p)}{c_p},
\label{eq:TLTP-A-2}
\ee
where $\Box_{\mu \nu}(p) \equiv p^2 \delta_{\mu \nu} - p_\mu p_\nu$ is the
usual transverse kinetic term and $c_p \equiv c(p^2/\Lambda^2)$ is a
dimensionless, smooth cutoff function. Substituting~\eq{eq:TLTP-A-2}
into~\eq{eq:TLTP-A}, noting that $\Box_{\mu \alpha}(p) \Box_{\alpha \nu}(p)
=p^2 \Box_{\mu \nu}(p)$ and integrating up we obtain
\[
    c(p) = p^2 \EP{AA}(p),
\]
where we have set the integration constant to zero to
ensure that the effective propagator is well behaved
as $p\rightarrow \infty$~\cite{YM-1-loop}.
Thus, in the $A$-sector, the effective
propagator relation takes the form
\be
    S^{\ AA}_{0 \mu\, \nu} (p) \EP{AA}(p) = \delta_{\mu \nu} - \frac{p_\mu p_\nu}{p^2};
\label{eq:EP-A}
\ee
in other words, the effective propagator is only the inverse of the two-point,
tree level vertex in the transverse space.

We have purposely left the $\psi$ sector until last, since
there is a subtlety here. The point is that since $\psi$
is a massless fermion (and we do not require the kinetic
term to be regulated by a cutoff function), we can choose
the seed action such that $S^{\ \psibar \psi}_0(p)$ is
actually independent of $\Lambda$. In this case, we choose
the integration constant in the $\psi$-sector version
of~\eq{eq:TLTP-chi} such that we have the relationship
\be
    S^{\ \, \psibar \psi}_{0 \, a \, c}(p) \EPB{\psibar \psi}{c \, b}(p)  = \delta_{ab}.
\label{eq:EP-psi}
\ee

The three equations~\eq{eq:EP-chi}, \eq{eq:EP-A} and~\eq{eq:EP-psi}
can be combined into
\be
    S_{0 MR}^{\ \, X \, Y}(p) \EPB{Y Z}{RN}(p) = \delta_{MN} - p'_M p_N
\label{eq:EffPropReln}
\ee
where we use the following prescription for interpreting
the various elements of this equation in different sectors:
\begin{enumerate}
    \item   in the $A$-sector, the indices are all Lorentz indices,
            $\EPB{Y Z}{RN}(p)$ is just $\EP{YZ}(p)\delta_{\rho \nu}$ and
            $p'_M = p_\mu / p^2$ and $p_N = p_\nu$;

    \item   in the $\psi$ and $\chi$-sectors, all indices
            are spinor indices but $p'_M$ and $p_N$ are null.
\end{enumerate}

We call the structure $p'_M p_N$ a `gauge remainder'. This captures
the notion that the pure gauge effective propagator leaves something
behind, other than a Kronecker-$\delta$, when contracted into a two-point,
tree level vertex. Equation~\eq{eq:EffPropReln}  is represented diagrammatically
in figure~\ref{fig:EP}, where  a dummy effective propagator is denoted by a long, solid
line.
\bcf[h]
    \be
        \cd{EffPropReln} \equiv \delta_{MN} - p'_M p_N  \equiv \delta_{MN} - \cd{FullGaugeRemainder}
    \label{eq:EffPProp}
    \ee
\caption{The effective propagator relation.}
\label{fig:EP}
\ecf

We conclude this section by giving explicit algebraic realisations
(that correspond to the regularisation described in appendix~\ref{app:Reg})
for the two-point, tree level vertices and effective propagators
which are summarised, together with the gauge remainders, in table~\ref{tab:TLTPs+EPs}.

\bct[h]
    \[
    \renewcommand{\arraystretch}{2}
    \begin{array}{c|cccc}
            & S_{0 MN}^{\ \,f\, f}(p)           & \EP{ff}(p)                    & p'_M                  & p_N
    \\[1ex] \hline
        A   & \ds \frac{\Box_{\mu \nu}(p)}{c_p} & \ds \frac{c_p}{p^2}           & \ds \frac{p_\mu}{p^2} & p_\nu
    \\[1ex]
        \psi& \psl                              & \ds \frac{1}{\psl}            &\mbox{---}             & \mbox{---}
    \\[2ex]
        \chi& \psl + \Lam                       & \ds \frac{1}{\psl + \Lam}     &\mbox{---}             & \mbox{---}
    \end{array}
    \renewcommand{\arraystretch}{1}
    \]
\caption{Algebraic realisation of the two-point, tree level vertices,
effective propagators and gauge remainders.}
\label{tab:TLTPs+EPs}
\ect

There are several things to note. First, the renormalisation condition~\eq{eq:RenormCondition}
demands that
\be
    c_0 = 1.
\label{eq:c_0}
\ee
Secondly, it is indeed the case that $\DEP{\psibar \psi}$ vanishes.

\section{Computing $\beta_1$}   \label{sec:beta1}

\subsection{The Starting Point}

The key to extracting $\beta$-function coefficients from
the weak coupling flow equations~\eq{eq:WeakCouplingExpansion-Flow-B} is to
use the renormalisation condition~\eq{eq:RenormCondition},
which places a constraint on  the vertex
$S^{AA}_{\mu \, \nu}(p)$. From equations~\eq{eq:WeakCouplingExpansion-Action} and~\eq{eq:c_0}
and table~\ref{tab:TLTPs+EPs}, this constraint is saturated at tree level:
\[
S^{A A}_{\mu \, \nu}(p) = \frac{1}{e^2}\Box_{\mu \nu}(p) + \Op{4} = \frac{1}{e^2} \frac{\Box_{\mu \nu}(p)}{c_0} + \Op{4} =  \frac{1}{e^2} S^{\ AA}_{0 \mu\, \nu}(p).
\]
Hence, all higher loop two-point vertices, $S_{n\geq 1 \mu \, \nu}^{\ \ \ \ AA}(p)$ vanish at $\Op{2}$.

To utilise this information, we
specialise equation~\eq{eq:WeakCouplingExpansion-Flow-B} to
compute the flow of $S_{1 \mu \,\nu}^{\ AA}(p)$:
\be
    \dot{S}_{1 \mu\, \nu}^{\ AA}(p) = -2\beta_1 \TLTP{p}
    + \sum_{r=0}^1 a_0[\bar{S}_{1-r},\bar{S}_r]^{AA}_{\mu \,\nu}(p) -  a_1[\Sigma_0]^{AA}_{\mu \, \nu}(p).
\label{eq:Beta1-Defining-pre}
\ee

Now we focus on the $\Op{2}$ part of this equation. The term on the \lhs, being
a two-point gauge vertex of loop order greater than zero, vanishes at $\Op{2}$.
On the \rhs, the
$a_0$ term can be discarded. Recall that this term comprises two vertices joined together
by an \ERG\ kernel (\cf\ figure~\ref{fig:FlowEx}). We must decorate each of these vertices with one
of the external fields, $A_\mu$ or $A_\nu$, else one of vertices is
one-point vertex and
these do not exist. However, if we decorate both vertices with a single
external field then, by gauge invariance, they are both transverse in $p$
and so the diagram is at least $\Op{4}$.

Equation~\eq{eq:Beta1-Defining-pre} collapses to an algebraic expression for
$\beta_1$:
\be
2 \beta_1 \Box_{\mu \nu}(p) + \Op{4} = -a_1[\Sigma_0]^{AA}_{\mu \, \nu}(p),
\label{eq:Beta1-Defining}
\ee
which is shown diagrammatically in figure~\ref{fig:b1:L0},
employing the notation of figure~\ref{fig:Diags:FE}.
It is taken as understood in all that follows that
the external indices are $\mu$ and $\nu$ and
we are working at $\Op{2}$.
\bcf[h]
    \be
        2 \beta_1 \Box_{\mu \nu}(p) + \Op{4} = -\frac{1}{2} \dec{\cd{Padlock-Sig_0}}{AA}
        =
        -\frac{1}{2}
        \dec{
            \begin{array}{ccc}
                \PD{L0-W}{fig:L0-W-Isolate} &   & \CD{L0-S}{L2-0h}
            \\[1ex]
                \cd{Padlock-S_0}            &-2 & \cd{Padlock-hS_0}
            \end{array}
        }{AA}
    \label{eq:b1:L0}
    \ee
\caption{A diagrammatic representation of the equation for $\beta_1$.
On the \rhs, we implicitly take the indices to be $\mu$ and $\nu$ and
work at $\Op{2}$.}
\label{fig:b1:L0}
\ecf

Diagrams are labelled in boldface. If a diagram is cancelled, then
its reference number is enclosed in curly braces, together with
the reference number of the diagram against which it cancels.
If the reference number of a diagram
is followed by an arrow, it can mean one of two things:
\begin{enumerate}
    \item   $\rightarrow 0$ denotes that the corresponding diagram can be set to zero, for some reason;

    \item   $\rightarrow$ followed by a number (other than zero)
            indicates the number of the figure in which the corresponding
            diagram is processed.
\end{enumerate}

\subsection{Diagrammatic Manipulations}

As it stands, we cannot directly extract a value for $\beta_1$
from equation~\eq{eq:b1:L0}. The \rhs\ is phrased in terms of
non-universal objects. Whilst one approach would be to choose a
particular scheme in which to compute these
objects~\cite{GI-ERG-II,Thesis} we anticipate
from~\cite{YM-1-loop,Thesis,scalar2loop} that this is unnecessary:
owing to the universality of $\beta_1$, all non-universalities
must somehow cancel out. To proceed, we utilise the flow
equations.

Our aim is to try and reduce the expression for $\beta_1$
to a set of $\Lambda$-derivative terms---terms where
the entire diagram is hit by $\flow$---since,
as we will see in section~\ref{sec:b1:Numerics},
such terms either vanish directly or give
universal contributions (in the limit that $D \rightarrow 4$).
Focusing on diagram~\ref{L0-W}, we note that if both
decorative fields decorate the vertex, then the
internal line is just a differentiated effective propagator---in
this case we can generate a
$\Lambda$-derivative term by moving the $\flow$
from the effective propagator to the vertex.

Hence, our first task is to separate off the manipulable component
of diagram~\ref{L0-W}. To this end, we define an object called
the \emph{reduced} kernel:
\be
    \left(\dot{\Delta}\right)_R \equiv \stackrel{\circ}{\Delta} =  \dot{\Delta} - \dot{\Delta}^{XY}_{ST}(k),
\label{eq:RW}
\ee
where we have suppressed all arguments of the generic kernel, $\dot{\Delta}$, and its reduction, $\stackrel{\circ}{\Delta}$.
In figure~\ref{fig:L0-W-Isolate} we re-express diagram~\ref{L0-W}.
\bcf[h]
    \[
        - \frac{1}{2} \dec{\cd{Padlock-S_0}}{AA} \equiv
        - \frac{1}{2}
        \dec{
            \begin{array}{ccc}
                \PD{S_0-DEP}{fig:b1:L1} &   & \CD{S_0-RW}{L2-S_0-RW-b}
            \\[1ex]
                \cd{S-0-DEP}            & + & \cd{S-0-RW}
            \end{array}
        }{AA}
    \]
\caption{Isolating the manipulable component of diagrams~\ref{L0-W}.}
\label{fig:L0-W-Isolate}
\ecf

The symbol $\DAD$ tells us that the corresponding kernel is not decorated,
whereas the symbol $\circ$ means the converse. We now convert diagram~\ref{S_0-DEP}
into a $\Lambda$-derivative term, as shown in figure~\ref{fig:b1:L1}.

\bcf[h]
    \beas
    - \frac{1}{2} \dec{\cd{S-0-DEP}}{AA}
    & = &
        -\frac{1}{2}
        \dec{
            \dec{\cd{Vertex-S_0-EP}}{\bullet} - \cd{Vertex-S_0-LdL-EP}
        }{AA}
    \\[1ex]
    &\equiv&
        -\frac{1}{2}
        \dec{
            \begin{array}{ccc}
                \PD{L1-LdL}{fig:Summary}        &   & \PD{L1-C}{fig:b1:L2}
            \\[1ex]
                \dec{\cd{Vertex-S_0}}{\bullet}  & - & \cd{Vertex-S_0-LdL}
            \end{array}
        }{AA\Delta}
    \eeas
\caption{The manipulation of diagram~\ref{S_0-DEP}.}
\label{fig:b1:L1}
\ecf

Notice that, on the second line, we have refined the notation further,
by promoting the effective propagator in both diagrams to an
implicit decoration. This must be done with care.
In diagram~\ref{L1-LdL},
the vertex is enclosed in square brackets which
tells us that $\flow$ is take to act \emph{after explicit decoration}.
However, in diagram~\ref{L1-C}, it is just the vertex which is
struck by $\flow$.
Diagram~\ref{L1-LdL} is one of the $\Lambda$-derivative terms
we have been looking for.

The next step is to process diagram~\ref{L1-C}, using the tree level
flow equations. This is shown in figure~\ref{fig:b1:L2}.
\bcf[h]
    \[
    \frac{1}{2}
    \dec{
        \cd{Vertex-S_0-LdL}
    }{AA \Delta}
    =
    \frac{1}{4}
    \dec{
        \PDi{Dumbbell-S_0b-S_0b}{L1-Db-S_0hx2}{fig:L1:Isolate}
    }{AA \Delta}
    \]
\caption{The manipulation of diagram~\ref{L1-C}.}
\label{fig:b1:L2}
\ecf

A word is in order about the rules for decorating diagram~\ref{L1-Db-S_0hx2}~\cite{Thesis}.
Let us start with the external fields. If we attach each of these to
a different structure, we pick up a factor of two; if we attach them to the
same structure, the combinatoric factor is just unity. Similarly, if the
ends of the effective propagator attach to different structures then we
pick up a factor of two; if they do not then the combinatoric factor is unity.

To proceed further, we now isolate all two-point, tree level
vertices (there was no need to do this when manipulating diagram~\ref{L0-W},
since the vertex of this diagram is compelled to be four-point). The reason
we do this
is that we do not wish to process such vertices via the flow equations;
rather, if we attach them to effective propagators, then we can use the
effective propagator relation. Alternatively, if we decorate them
with external fields, we will be able to use the fact that they are manifestly $\Op{2}$.

To facilitate this separation, we define reduced, tree level vertices, thus:
\[
    v_0^R = v_0 - v_{0 RS}^{\ f \, f}(k),
\]
where we have suppressed all arguments of the generic vertex $v_0$ and its reduction, $v_0^R$.
We now re-express diagram~\ref{L1-Db-S_0hx2}, as in figure~\ref{fig:L1:Isolate}, recalling the
notation of figure~\ref{fig:SpecificNumDecs}. Notice that, for those diagrams possessing
two-point, tree-level, vertices, we have expanded out the bar-notation according
to~\eq{eq:NFE:BarNotation}:
\[
    a_0[\bar{S}_0, \bar{S}_0^2] = a_0[S_0, S_0^2] -  a_0[S_0, \hS_{0}^2] - a_0[\hat{S}_0, S_0^2] = - a_0[\hat{S}_0, S_0^2],
\]
where the cancellation of terms occurs due to the equality of the Wilsonian effective action and seed
action two-point, tree level vertices.
\bcf[h]
    \[
    \frac{1}{4}
    \dec{
        \cd{Dumbbell-S_0b-S_0b}
    }{AA\Delta}
    =
    \frac{1}{4}
    \dec{
        \begin{array}{ccccc}
            \PD{L1-Db-S_0Rx2}{fig:L1:IsolateManip}  &   & \PD{L1-Db-S_0R-TLTP}{fig:L1:PartialDec}   &   & \PD{L1-Db-TLTPx2}{fig:L1:PartialDec}
        \\[1ex]
            \cd{Dumbbell-S_0bR-S_0bR}               &-2 & \cd{Dumbbell-S_0hR-TLTP}                  & - & \cd{Dumbbell-TLTPx2}
        \end{array}
    }{AA\Delta}
    \]
\caption{Isolating the two-point, tree level vertices of diagram~\ref{L1-Db-S_0hx2}.}
\label{fig:L1:Isolate}
\ecf

Our strategy now is to decorate the two-point, tree level vertices
of diagrams~\ref{L1-Db-S_0R-TLTP} and~\ref{L1-Db-TLTPx2}. For each such
vertex we can attach one of two things: either an end of the effective
propagator or an external field. In the former case, we must now
join up the loose end of the effective propagator. This can never attach
to the kernel, as we now argue. The key point is that
only $\chi$ or $\psi$-sector kernels have decorations
and that these decorations must be $A$s. Thus, trying to join one end
of the effective propagator to the kernel and the other end to a
two-point, tree level vertex would mean that the vertex would have to be
decorated by an $A$ and one of the spinor fields; such vertices do not
exist. Consequently, having attached one end of the effective propagator
to a two-point, tree level vertex, the other end must attach to the other vertex.

We can attach an external field to the two-point, tree level vertex of
diagram~\ref{L1-Db-S_0R-TLTP} but we cannot do likewise in
diagram~\ref{L1-Db-TLTPx2}. Performing this decoration forces the kernel to
be in the $A$-sector; hence the kernel cannot be decorated. But, in
diagram~\ref{L1-Db-TLTPx2}, this would mean that it is not
possible to perform the explicit decorations, as there are too many fields
and not enough legal locations.

The result of the partial decoration of diagrams~\ref{L1-Db-S_0R-TLTP}
and~\ref{L1-Db-TLTPx2} is
shown in figure~\ref{fig:L1:PartialDec}.
\bcf[h]
    \[
    \begin{array}{r}
        \ds
        -\frac{1}{2}
        \dec{
            \cd{Dumbbell-S_0hR-TLTP} +  \hf \cd{Dumbbell-TLTPx2}
        }{AA\Delta}
        =
        -
        \dec{
            \begin{array}{ccc}
                \PD{L1-EPReln}{fig:L1:EffProp}  &           & \PD{L1-EPRelnB}{fig:L1:EffProp}
            \\[1ex]
                \cd{Struc-0hR-W-TLTP-EP}        & + \ds \hf & \cd{Struc-TLTP-W-TLTP-EP}
            \end{array}
        }{AA}
    \\[10ex]
        \ds
        -\hf
        \dec{\PDi{Struc-TLTP-E-W-0hR}{L1-E}{fig:Summary}}{A\Delta}
    \end{array}
    \]
\caption{Result of decorating the two-point, tree level vertices of diagrams~\ref{L1-Db-S_0R-TLTP}
and~\ref{L1-Db-TLTPx2}.}
\label{fig:L1:PartialDec}
\ecf

We now reach the crux of the entire diagrammatic approach: diagrams~\ref{L1-EPReln}
and~\ref{L1-EPRelnB}
can be processed using the effective propagator relation~\eq{eq:EffPProp} upon which,
the calculation starts to simplify. This is shown in figure~\ref{fig:L1:EffProp}.
\bcf[h]
    \[
        \begin{array}{c}
            -\dec{\cd{Struc-0hR-W-TLTP-EP}+ \ds \hf \cd{Struc-TLTP-W-TLTP-EP}}{AA}
        \\[10ex]
            =
            -
            \dec{
                \begin{array}{ccccccc}
                        \PD{L1-hS0-W}{fig:L2:combine}   &   & \Discard{L1-hS0-W-GR} &       &   \PD{L1-TLTP-W}{fig:L2:combine}  &       &\Discard{L1-TLTP-W-GR}
                \\[1ex]
                        \cd{Padlock-hS_0R}              & - & \cd{Padlock-hS_0R-GR} &+\ds\hf& \cd{Padlock-TLTP}                 &-\ds\hf&\cd{Padlock-TLTP-GR}
                \end{array}
            }{AA}
        \end{array}
    \]
\caption{Applying the effective propagator relation to diagram~\ref{L1-EPReln}.}
\label{fig:L1:EffProp}
\ecf

%\CancelCom{L1-hS0-W}{L0-S}{: the decorations force the vertex of diagram~\ref{L0-S}
%to be four-point, and thus reduced.}

Diagram~\ref{L1-hS0-W-GR} can be discarded: for the gauge remainder to have
support, it must be in the $A$-sector (see table~\ref{tab:TLTPs+EPs}). However,
this means that the vertex is pure $A$ and so is killed by the gauge
remainder, courtesy of the Ward identity~\eq{eq:WI-A}. Diagram~\ref{L1-TLTP-W-GR}
can be discarded: the kernel is decorated, meaning that it must be in
either the $\psi$ or $\chi$-sectors and so the gauge remainder has no support.

Our strategy now is simple: we iterate the diagrammatic procedure. We begin
by taking diagram~\ref{L1-Db-S_0Rx2} and isolating the manipulable
component \ie\ the component which possesses only Wilsonian effective action
vertices and an undecorated kernel. This is shown in figure~\ref{fig:L1:IsolateManip}.
\bcf[h]
    \[
    \frac{1}{4} \dec{\cd{Dumbbell-S_0bR-S_0bR}}{AA\Delta} =
    \frac{1}{4}
    \dec{
        \begin{array}{ccccc}
            \PD{L0-Db-M}{fig:L2-LdL}    &   &\Discard{L1-Db-NM}             &   & \CD{L0-Db-S}{L2-S}
        \\[1ex]
            \cd{Dumbbell-S_0R-S_0R-DEP} & + & \cd{Dumbbell-S_0R-S_0R-RW}& -2& \cd{Dumbbell-S_0hR-S_0R}
        \end{array}
    }{AA\Delta}
    \]
\caption{Isolating the manipulable component of diagram~\ref{L1-Db-S_0Rx2}.}
\label{fig:L1:IsolateManip}
\ecf

Diagram~\ref{L1-Db-NM} can be discarded. Since one-point vertices do not
exist, and the vertices cannot be two-point (by the definition of a
reduced vertex), both vertices must be at least three-point. But, upon
explicit decoration, this does not leave behind any fields to decorate
the (reduced) kernel.

In figure~\ref{fig:L2-LdL}, we convert diagram~\ref{L0-Db-M} into a
$\Lambda$-derivative term.
\bcf[h]
    \[
        \frac{1}{4}
        \dec{\cd{Dumbbell-S_0R-S_0R-DEP}}{AA\Delta}
        =
        \frac{1}{16}
        \dec{
            \begin{array}{ccc}
                \PD{L2-LdL}{fig:Summary}                        &   &\PD{L2-C}{fig:L2:M}
            \\[1ex]
                \dec{\scd[3]{Vertex-OR}{Vertex-OR}}{\bullet}    & -2&\dec{\scd[3]{Vertex-OR-LdL}{Vertex-OR}}{}
            \end{array}
        }{AA\Delta^2}
        -\frac{1}{4}
        \dec{
            \DO{\scd[3]{Vertex-OR-DEP}{Vertex-OR}}{L2-WC}
        }{AA\Delta}
    \]
\caption{Converting diagram~\ref{L0-Db-M} into a
$\Lambda$-derivative term.}
\label{fig:L2-LdL}
\ecf

The overall factors in front of the daughter diagrams
require explanation~\cite{Thesis}.
In going from the
parent diagram, \ref{L0-Db-M}, to the daughters, we have moved the
$\flow$ off the effective propagator. This effective propagator has
subsequently been promoted to an implicit decoration. Immediately, we
pick up a factor of $1/2$ in recognition of the fact that the
effective propagator can be reattached, so as to join the two vertices
back together, either way round.
We now have two, identical effective propagators amongst our
implicit decorations. To recreate the parent diagram, we
can choose either of these effective propagators to be hit by $\flow$
and so we must compensate with a further factor of $1/2$. This
explains why diagram~\ref{L2-LdL} comes with a relative factor
of $1/4$, compared to the parent. As for diagram~\ref{L2-C} the
$\flow$ can strike either of the vertices, with the same effect;
we combine these terms, to yield a factor of two.

Let us now consider diagram~\ref{L2-WC}, the final correction
generated when we convert the parent diagram into the $\Lambda$-derivative
term, \ref{L2-LdL}. Diagram~\ref{L2-WC} can be formed in four ways:
the $\flow$ can strike either of the effective propagators, and we can form
a loop with the differentiated effective propagator on either of the vertices.
Diagram~\ref{L2-WC} can be discarded. To make a complete diagram,
we must join the two vertices together, using the effective propagator, which
must be in the $A$-sector. Since the bottom vertex is reduced, and one-point
vertices do not exist, it must also be decorated by both of the external
fields.
Hence, the bottom vertex is a pure gauge, three-point vertex,
which vanishes as a consequence of charge conjugation
invariance (equivalently, Furry's theorem~\cite{furry}).

Now we process diagram~\ref{L2-C}. To do this, we must understand
how to compute the flow of a reduced vertex. This is easy. Remember
that a reduced vertex does not have a two-point, tree level
term. From section~\ref{sec:EffProp}, we know that the flow of a two-point,
tree level vertex produces a pair of two-point, tree level
vertices joined together by a differentiated effective propagator.
This diagram must be excluded from the flow of a reduced vertex,
meaning that either at least one of the vertices must be reduced
or the kernel must be decorated.
In figure~\ref{fig:L2:M} we show the result of processing
diagram~\ref{L2-C}.
\bcf[h]
    \[
    -\frac{1}{8}\dec{\scd[3]{Vertex-OR-LdL}{Vertex-OR}}{AA\Delta^2} =
    -\frac{1}{16}
    \dec{
        \begin{array}{ccccc}
            \Discard{L2-TooBig}                         &   & \PD{L2:Attach}{fig:L2:Partial}            &   & \PD{L2:Attach-B}{fig:L2:Partial}
        \\[1ex]
            \scd[6]{Dumbbell-S_0bR-S_0bR}{Vertex-OR}    &-2 & \scd[6]{Dumbbell-S_0hR-TLTP}{Vertex-OR}   & - & \scd[6]{Dumbbell-TLTPx2-RW}{Vertex-OR}
        \end{array}
    }{AA\Delta^2}
    \]
\caption{The result of processing diagram~\ref{L2-C}.}
\label{fig:L2:M}
\ecf

Diagram~\ref{L2-TooBig} can be discarded. All three vertices of this diagram are
reduced and there are not enough implicit decorations to avoid one of them being
either one-point, which does not exist, or two-point, which is forbidden by
the reduction.

Now we decorate the two-point, tree level vertex of diagram~\ref{L2:Attach}.
When doing so with an  effective propagator, we pick up a factor of four:
one factor of two to recognise that we could have attached either
effective propagator and one factor of two because the ends of the
effective propagator attach to different structures.
This time, we utilise the effective propagator relation immediately and again
recognise that all gauge remainders can be discarded.

Finally, we decorate the two-point, tree level vertices of diagram~\ref{L2:Attach-B}.
Since the kernel must be decorated---and so must be in the $\psi$ or $\chi$-sectors---we
cannot decorate either of the
two-point, tree level vertices with an external field. Consequently, we must
either attach a separate effective propagator to each or we must join
the two vertices together with a single effective propagator.
Again, we utilise the effective propagator relation immediately and
recognise that all gauge remainders can be discarded.
We thus obtain the
diagrams of figure~\ref{fig:L2:Partial}.

\bcf[h]
    \[
        \begin{array}{c}
            \ds
            \frac{1}{8}
            \dec{
                \scd[6]{Dumbbell-S_0hR-TLTP}{Vertex-OR} + \frac{1}{2} \scd[6]{Dumbbell-TLTPx2-RW}{Vertex-OR}
            }{AA\Delta^2}
            =
            \frac{1}{8}
            \dec{
                \PO{\scd[8]{Dumbbell-S_0hR-TLTP-E}{Vertex-OR}}{L2-E}{fig:Summary}
            }{A\Delta^2}
        \\[10ex]
            \ds
            +\frac{1}{2}
            \dec{
                \begin{array}{ccccccc}
                    \CD{L2-S}{L0-Db-S}      &   & \Discard{L2-S-Loop}               &       &\Discard{L2-Loop}                  &   &\PD{L2-S_0-RW}{fig:L2:combine}
                \\[1ex]
                    \cd{Dumbbell-S_0hR-S_0R}& + & \scd[3]{Vertex-OhR-DEP}{Vertex-OR}&+\ds\hf&\scd[3]{Vertex-TLTP-RW}{Vertex-OR} & + & \cd{S_0R-RW}
                \end{array}
            }{AA\Delta}
        \end{array}
    \]
\caption{Result of decorating the two-point, tree level vertices of diagrams~\ref{L2:Attach}
and~\ref{L2:Attach-B}.}
\label{fig:L2:Partial}
\ecf

\Cancel{L2-S}{L0-Db-S}

Diagrams~\ref{L2-S-Loop} and~\ref{L2-Loop} can be discarded for exactly the same
reason as diagram~\ref{L2-WC}. The final diagrammatic step is to combine
diagrams~\ref{L1-hS0-W}, \ref{L1-TLTP-W} and~\ref{L2-S_0-RW},
as shown in figure~\ref{fig:L2:combine}. We utilise the fact that
the two-point, tree level, Wilsonian effective action vertices are
equal to the corresponding seed action vertices and also that those
diagrams in which the vertex is two-point necessarily have a reduced
(\ie\ decorated) kernel.
\bcf[h]
    \[
        \dec{-\cd{Padlock-hS_0R} -\ds \hf \cd{Padlock-TLTP} + \hf \cd{S_0R-RW} }{AA} =
        -\dec{
            \begin{array}{ccc}
                \CD{L2-0h}{L0-S}    &       &\CD{L2-S_0-RW-b}{S_0-RW}
            \\[1ex]
                \cd{Padlock-hS_0}   &-\ds\hf& \cd{S-0-RW}
            \end{array}
        }{AA}
    \]
\caption{Result of combining diagrams~\ref{L1-hS0-W}, \ref{L1-TLTP-W} and~\ref{L2-S_0-RW}.}
\label{fig:L2:combine}
\ecf

\Cancel{L2-0h}{L0-S}
\Cancel{L2-S_0-RW-b}{S_0-RW}

The only surviving terms are diagrams~\ref{L1-LdL}, \ref{L1-E}, \ref{L2-LdL} and~\ref{L2-E}.
We now explicitly decorate these diagrams and, throwing away any terms which vanish
due to charge conjugation invariance, arrive at figure~\ref{fig:Summary}.
\bcf[h]
    \[
    2\beta_1 \Box_{\mu \nu} (p) + \Op{4} =
    -\frac{1}{2}
    \dec{
        \begin{array}{ccc}
            \LD{Beta1-LdL-A}    &   & \LD{Beta1-LdL-B}
        \\[1ex]
            \cd{Beta1-LdL-A}    & - & \cd{Beta1-LdL-B}
        \end{array}
    }{\bullet}
    -
    \dec{
        \begin{array}{cccc}
            \LDLB{Beta1-c0-A}   &   & \LDRB{Beta1-c0-B} & \hspace{-1em} \scriptstyle \rightarrow 0
        \\[1ex]
            \cd{Beta1-c0-A}     & - & \cd{Beta1-c0-B}   &
        \end{array}
    }{}
    \]
\caption{The surviving contributions to $\beta_1$.}
\label{fig:Summary}
\ecf

A number of comments are in order. All fields in diagram~\ref{Beta1-LdL-A}
attach to the same vertex and so we do not pick up any factors upon
decoration. In diagram~\ref{Beta1-LdL-B}, each of the effective propagators
can be attached either way round, giving a factor of four. The external
fields attach to different vertices, yielding a further factor of two.

In diagrams~\ref{Beta1-c0-A} and~\ref{Beta1-c0-B} we recognise that
the internal field leaving the two-point, tree level vertex must be
in the $A$-sector. Let us now analyse these diagrams in more detail.
The two-point, tree level vertex is, from table~\ref{tab:TLTPs+EPs},
at least $\Op{2}$ and so, since we are working at $\Op{2}$, we are
compelled to take $\Op{0}$ from the rest of each of the diagrams.
This is no problem for the differentiated effective propagator,
which will contribute $\sim c'_0$. However, we can
straightforwardly demonstrate~\cite{Thesis}
using the Ward identities~\eq{eq:WI-A} and~\eq{eq:WI-F}
and the effective propagator relation, that the sum of the
diagrams to which the top end of the effective propagator
attaches are transverse in $p$---and hence at least $\Op{2}$.\footnote{This
can be seen by contracting the four point vertex of diagram~\ref{Beta1-c0-A}
with the momentum of one of the external fields and likewise
for the top-most three-point vertex of diagram~\ref{Beta1-c0-B}. In the latter
case, this procedure generates two-point vertices to which we can apply
the effective propagator relation.}
Thus, the sum of diagrams~\ref{Beta1-c0-A} and~\ref{Beta1-c0-B}
is at least $\Op{4}$ and so does not contribute to $\beta_1$.
That this is the case is just as well: $c'_0$ is non-universal
and its inverse cannot be generated by loop integration.

\subsection{The $\Lambda$-derivatives}  \label{sec:b1:Numerics}

\subsubsection{Strategy}

From figure~\ref{fig:Summary}, our equation for $\beta_1$ can
be written in the form:
\[
    2\beta_1 \Box_{\mu \nu} (p) = -\frac{1}{2} \dec{\OLDs}{\bullet}.
\]
We now want to make the integral over loop momentum (which
we will take to be $k$) to be explicit and so write
\be
    2 \beta_1 \Box_{\mu \nu} (p) = -\frac{1}{2} \int_k \dec{\OLDs(k)}{\bullet}.
\label{eq:TDM:Interchange}
\ee

The next step that we wish to perform is to interchange the order of the $\Lambda$-derivative
and the momentum integral. This step is trivial only if the integral is convergent, even after this
change.
We temporarily ignore this subtlety and so now have
\[
    2 \beta_1 \Box_{\mu \nu} (p) = -\frac{1}{2} \dec{\int_k \OLDs(k)}{\bullet}.
\]

Since the \lhs\ of this equation comprises a number times $\Op{2}$, it follows
that the coefficient multiplying the $\Op{2}$ part of the \rhs\ must be
dimensionless. Consequently, we can schematically write
\[
    \beta_1 = \flow \left( \mbox{Dimensionless Quantity} \right).
\]
For the \rhs\ to survive differentiation \wrt\ $\Lambda$,
it must either depend on some dimensionless running
coupling---other than $e$---or there must be some scale, other than $\Lambda$,
available for the construction of dimensionless quantities.
We show in appendix~\ref{app:Running} that no such running couplings
exist.

One scale which is available is $p$ and so we can envisage contributions to $\beta_1$ of
the form (in $D=4$)
\[
    \flow \ln p^2 / \Lambda^2.
\]
Indeed, one can arrange the calculation precisely so as to obtain just such a contribution.
However, there is an easier way to proceed. We note that the \lhs\ of
equation~\eq{eq:TDM:Interchange} is Taylor expandable in $p$ and so
it must be the case that the \rhs\ is Taylor expandable too.
Let us suppose that we were to Taylor expand $\OLDs$ in equation~\eq{eq:TDM:Interchange}
\ie\ \emph{before} we have interchanged the order of loop integration and
differentiation \wrt\ $\Lambda$.

If we now try and change the order of loop integration and
differentiation \wrt\ $\Lambda$ we must be very careful,
since this procedure has the capacity to introduce
divergences in both the IR and UV---we comment further on
this shortly.
Thus, to legally move $\flow$ outside of the loop integral
in the case where we have Taylor expanded in $p$, we must introduce some regulator.
This then provides the scale necessary to form dimensionless quantities. After allowing
$\flow$ to act, this unphysical scale will disappear. As we are already working in
dimension $D$, it is natural to use dimensional regularisation as our
regulator.

We are now able to see the effects of keeping diagrams
generated by the $\psi$-sector terms in the flow equation
(recall that, since $\DEP{\psibar \psi}=0$, these could have
been discarded). If we keep them then, after first Taylor expanding in $p$
and next interchanging
the order of loop integration and differentiation,
$\int_k \OLDs(k)$ is still convergent in the UV but diverges in the
IR. Had we discarded the $\psi$-sector diagrams, then $\int_k \OLDs(k)$
would diverge in the UV but converge in the IR. Either way, the contribution
to $\beta_1$ in the $D \rightarrow 4$ is the same, but we have a choice
about how to compute it.

For the purposes of \QED, it makes no difference whether we
extract $\beta_1$ from the IR or the UV of $\int_k \OLDs(k)$
(though, in the former case, it makes it clear how universal
contributions are controlled by the renormalisation condition).
However, in Yang-Mills theory, there are only a subset of diagrams
for which we have this choice; the rest yield contributions from
the IR only~\cite{YM-1-loop,Thesis}. Consequently, in Yang-Mills,
it makes sense to evaluate all contributions to $\beta$-function
coefficients in the IR. As the purpose of this paper is to
introduce and illustrate the Yang-Mills methodology, we will
choose to compute in the IR here, too.

With this in mind, our strategy for extracting $\beta_1$ is as follows.
First,
we Taylor expand all diagrams to $\Op{2}$. Then we focus
on the IR divergent part of the integral, throwing all
other contributions away (since these will vanish in
the limit that $D \rightarrow 4$).

\subsubsection{Numerical Evaluation}    \label{sec:Numerics}

Let us start by looking at diagram~\ref{Beta1-LdL-A}.
This has a single effective propagator which,
scanning through table~\ref{tab:TLTPs+EPs},
will blow up when $p \rightarrow 0$ in the
$\psi$ or $A$-sectors.
However, this is not enough to generate an IR divergence,
and so the diagram vanishes in the $D\rightarrow 4$ limit.

Finally, let us look at diagram~\ref{Beta1-LdL-B}.
In the pure gauge sector, this diagram does not
even exist, since three-point pure gauge vertices
vanish. The diagram does exist in the $\psi$ and
$\chi$-sectors.
The severe IR behaviour
occurs in the $\psi$-sector and so we expect
to be able to throw the $\chi$-sector diagram away.
We can do this, but must be careful: the $\chi$-sector
diagram regulates the $\psi$-sector diagram in the UV.
We can throw away the $\chi$-sector diagram if
we incorporate the effects of the UV regularisation.
This is done
by replacing the upper limit of the radial integral by $\Lambda$
for the $\psi$-sector diagram, which is
valid up to (non-universal) corrections which vanish as $D \rightarrow 4$~\cite{Thesis}.
We now rescale $k \rightarrow k/ \Lambda$, so that the
upper limit of the radial integral becomes unity, and the diagram
acquires an overall factor of $\Lambda^{-2\epsilon}$.

Recalling the Feynman rules of section~\ref{sec:Diagrammatics} for
diagrams possessing internal $\psi$s, diagram~\ref{Beta1-LdL-B}
reduces to
\[
    -\dec{\Lambda^{-2\epsilon}}{\bullet} \ODIntL{k}{0}{1} k^{D-1} \ODInt{\!\! \AngVol{D}} \;
    \tr
    \left[
        S_{0 \ \ \ \, \mu}^{\ \psibar \psi A} (-p-k,k,p) \frac{1}{\ksl} S_{0 \ \ \ \, \nu}^{\ \psibar \psi A}(-k, p+k,-p) \frac{1}{\psl + \ksl}
    \right]_{p^2},
\]
where the notation $\left.f\right|_{p^2}$ signifies that the $O(p^2)$ in
the series expansion of $f$ must be taken and $\AngVol{D}$ is the area of
the unit sphere in $D$ dimensions, divided by $(2\pi)^D$. The leading IR behaviour
comes from taking the $\Op{2}$ part of $1/(\psl + \ksl)$, and so we must set
$p=0$ in the vertices. Now, setting $p$ to zero in this manner allows
us to relate the three-point vertices to two-point vertices via
the Ward identity~\eq{eq:WI-F}.

Indeed, specialising \eq{eq:WI-F} to the
momenta $\{-k-\eta,k, \eta\}$ with $\eta$ infinitesimal yields
\be
    \eta^\mu S_{0 \ \ \ \, \mu}^{\ \psibar \psi A} (-k-\eta,k, \eta)
    = S_0^{\ \psibar \psi}(-k-\eta) - S_0^{\ \psibar \psi}(-k),
\ee
from which
one immediately obtains $S_{0 \ \ \ \mu}^{\ \psibar \psi A}(-k,k,0) = \demu^k \,
S_0^{\ \psibar \psi}(-k)$ by expanding to first order in $\eta$.

We therefore have that
\[
    2\beta_1 \Box_{\mu \nu}(p) = \dec{\Lambda^{-2\epsilon}}{\bullet}\ODIntL{k}{0}{1} k^{D-1} \ODInt{\!\! \AngVol{D}} \;
    \tr
    \left[
        \gamma_\mu \frac{\ksl}{k^2} \gamma_\nu
        \left[
            2 \psl \frac{k.p}{k^4} + \ksl \left( \frac{p^2}{k^2} - 4 \frac{(p.k)^2}{k^6} \right)
        \right]
    \right].
\]
Taking the trace of the $\gamma$ matrices and averaging over angles yields
\[
    2\beta_1 \Box_{\mu \nu}(p) =
    -4 \AngVol{D} \frac{D-2}{D+2} \dec{\Lambda^{-2\epsilon}}{\bullet}\ODIntL{k}{0}{1} k^{-1-2\epsilon} \left[p^2 \delta_{\mu\nu} - \frac{4}{D} p_\mu p_\nu \right],
\]
from which we obtain, in the $D \rightarrow 4$ limit,
\[
    \beta_1 = \frac{1}{12 \pi^2}.
\]

\appendix

\section{Regularisation}    \label{app:Reg}

In this appendix, we justify the statement that QED is regularised
by the introduction of the cutoff function $c$ in the photon
effective propagator and by the introduction of a massive
bosonic spinor $\chi$; the effective propagators are listed in
table~\ref{tab:TLTPs+EPs}.

It is easy to appreciate intuitively that the regularised theory can
be finite in $D=4$ dimensions, to all orders in perturbation
theory: in a given diagram generated by the flow, every loop
containing at least one photon effective propagator is regulated
by the presence of the cutoff function (providing $c$ decays
strongly enough, as determined below) whereas the ultraviolet
divergences in purely fermionic loops are cancelled by those
coming from purely commuting spinor loops.

In order to prove that the regularisation is sufficient for the
computations in this paper, we would need to further constrain
the, so far largely undetermined, seed action $\hS$. Rather
than doing this, we follow the spirit of refs.~\cite{Thesis,scalar2loop}, leaving
$\hS$ largely undetermined but insisting that the underlying
regularisation is sufficient to regularise the theory defined by
the simplest bare action:
\be \label{regsqed}
    S_{\rm bare} = \frac{1}{4 e^2} F_{\mu \nu} \kernel{\ci} F_{\mu \nu}
    + \frac{1}{e^2} \Int{x}
    \left[
        \psibar (i \desl + \asl ) \psi + \chibar (i \desl + \Lam + \asl ) \chi
    \right].
\ee
We will see that this is the case providing $c(x)$ decays faster
than $1/x^2$. Although we do not prove it, it is undoubtedly
true that the simplest $\hS$ then also leads to a fully
regularised theory. (A larger class of $\hS$ will also result
in regularised diagrams and by requiring $c$ to decay faster
even more general seed actions may be considered.)

Using \eq{regsqed}, the worst case divergence from a $\psi$ loop
is the one-loop photon self-energy, which is quadratically
divergent by power counting. However as is well known, this must be
transverse~\cite{itz}, see also equation~\eq{eq:WI-A}, and thus is only
logarithmically divergent. This divergence is exactly cancelled by
the opposite sign $\chi$ loop. The only other possible cause for
concern would be the one-loop four-photon vertex (vertices with
just an odd number of photons vanishing by Furry's theorem~\cite{furry})
but again, as is well known~\cite{itz},
gauge invariance forces this diagram to be finite.

It remains only to show that $c$ can be chosen to make all the
other diagrams ultraviolet finite. We follow standard power
counting techniques, adapting ref.~\cite{SU(N|N)}. Although we should gauge
fix for this, the ghosts play no \role, decoupling in both the
Lagrangian and the Ward identities.

The superficial degree of divergence of any
one-particle-irreducible diagram in $D$ space-time dimensions, is
given by
\be \label{sdd}
{\cal D}_{\Gamma} = D L - (2n+2) \, I_{A} - I_{\psi} - I_{\chi},
\ee
where $L$ is the number of loops and $I_{f}$ stands for the number
of internal lines of $f$-type. Note that the vertices from
\eq{regsqed} do not enter \eq{sdd} as they do not carry any
momentum dependence.

The variables which ${\cal D}_{\Gamma}$ depends upon can be easily
related to the number of external lines and vertices of each
type---respectively $E_{f}$, $V_f$---as
\bea
L &=& 1 + I_{A} + I_{\psi}
+ I_{\chi} - V_{{\psi}^2 A} - V_{{\chi}^2 A}, \label{lrel}\\ E_{A}
&=& -2 I_{A} + V_{{\psi}^2 A} + V_{{\chi}^2 A}, \label{earel}\\
E_{\psibar} &=& - I_{\psi} + V_{{\psi}^2 A}, \label{epsirel}\\
E_{\chibar} &=& - I_{\chi} + V_{{\chi}^2 A}. \label{echirel}
\eea
In equations~\eq{epsirel} and~\eq{echirel}, by $E_{\psibar,\chibar}
(= E_{\psi,\chi})$ we mean the number of external conjugate spinor
lines.

Using the above formulae, we rewrite \dga\ so that it is
independent of internal lines:
\be \label{sdd2}
{\cal D}_{\Gamma} = (D-2n-4) \, (L-2) - E_{A} -(2n+3) \,
E_{\psibar} -(2n+3) \, E_{\chibar} + 2 (D-n-2).
\ee

In order for every possible $L\ge2$ loop 1PI diagram to have a
negative \dga, one can impose all coefficients in \eq{sdd2} to be
negative and, thus, get sufficient conditions. This results in a
lower bound for $n$,
\be \label{finalcons}
n > D-2.
\ee
Such a condition is also necessary if one relies on power counting
arguments only, as $2 (D-n-2)$ represents the actual degree of
divergence of the two-loop vacuum diagram made from two
three-point vertices (the so-called no-go diagram).

For the one-loop diagrams we set $L=1$ and, rearranging some of its
terms, it is easy to show that diagrams containing at least one
external conjugate spinor and those with more than $D$ external
photons are finite provided $n$ satisfies \eq{finalcons}. On the
contrary, one-loop diagrams with up to $D$ external photons and no
external $\psibar$ or $\chibar$ cannot be regulated this way,
whatever the choice of $n$, since \dga\ is then $D-E_A$. In four
dimensions, these are just the one-loop two-point and four-point
photon vertices that we have already explained are finite.

We have thus shown that providing $n>2$, the theory defined by
\eq{regsqed} is finite to all orders in perturbation theory in
$D=4$ dimensions.

\section{Dimensionless Running Couplings}   \label{app:Running}

Consider the flow of any vertex with mass dimension $\geq 0$. Taylor
expand in momenta and focus on the term which is the same
order in momenta as the mass dimension of the vertex. The
coefficient of this term must be dimensionless. We now demonstrate
that the flow of all such coefficients vanishes. The only
candidates are as follows, where we indicate the power of
momentum that we must take from the vertex:
\[
    \left. S^{\psibar \psi}(k) \right|_{\mathrm{mom}^1},  \qquad
    \left. S^{\psibar \psi A}_{\ \ \ \, \mu}(k, p-k, p) \right|_{\mathrm{mom}^0}, \qquad
    \left. S^{AA}_{\mu \; \nu}(k)\right|_{\mathrm{mom}^2},  \qquad  \psi(\psibar) \rightarrow \chi(\chibar).
%   \left. S^{AAA}_{\mu \; \nu \, \rho}(k, p-k, p)\right|_{\mathrm{mom}^1}, \qquad
%   \left. S^{AAAA}_{\mu \; \nu \; \rho \, \sigma}(k, p-k, p-l,l)\right|_{\mathrm{mom}^0}.
\]
(Of the other potential candidates, $S^{AAA}$ does not exist by charge
conjugation invariance and $S^{AAAA}$ does not have an $\order{\mathrm{mom}^0}$
component since, by gauge invariance, it is transverse on all legs.)

The point is that all of these vertices, expanded to the order in momenta
indicated, are controlled by the renormalisation condition~\eq{eq:RenormCondition}
and so can immediately be shown to be independent of $\Lambda$. Therefore, there
are no dimensionless running coupling constants, besides $e$.

\paragraph{Acknowledgements}
TRM and OJR acknowledge financial support from PPARC Rolling Grant
PPA/G/O/2002/0468. TRM is grateful for support from a CERN paid
scientific associateship during this research. SA would
like to thank Massimo Testa and Kensuke Yoshida
for their enthusiasm and interest.


\begin{thebibliography}{105}

\bibitem{YM-1-loop}         S.~Arnone, A.~Gatti and T.~R.~Morris, \PhysRev{D}{67}{2003}{085003}, \arxiv{\hepth{0209162}}.
							%%CITATION = HEP-TH 0209162;%%


\bibitem{Quarks2004}        O.~J.~Rosten, T.~R.~Morris and S.~Arnone, The Gauge Invariant ERG,
                            Proceedings of Quarks 2004, Pushkinskie Gory, Russia, 
							24-30 May 2004, \http{quarks.inr.ac.ru},
                            \arxiv{\hepth{0409042}}.
							%%CITATION = HEP-TH 0409042;%%

\bibitem{Thesis}            O.~J.~Rosten, `The Manifestly Gauge Invariant Exact Renormalisation Group', Ph.D.\ Thesis,
							\arxiv{\hepth{0506162}};
                            S.~Arnone, T.~R.~Morris and O.~J.~Rosten, \arxiv{\hepth{0507154}}.
							%%CITATION = HEP-TH 0506162;%%
							%%CITATION = HEP-TH 0507154;%%

\bibitem{SU(N|N)}           S.~Arnone, Y.~A.~Kubyshin and T.~R.~Morris, J.~F.~Tighe,
                            \IntJModPhys{A}{17}{2002}{2283}, \arxiv{\hepth{0106258}}.
							%%CITATION = HEP-TH 0106258;%%

\bibitem{TRM+JL-0}          J.I.~Latorre and T.~R.~Morris, \jhep{0011}{2000}{004}, \arxiv{\hepth{0008123}}.
							%%CITATION = HEP-TH 0008123;%%

\bibitem{scalar2loop}       S.~Arnone, A.~Gatti,  T.~R.~Morris and O.~J.~Rosten, \PhysRev{D}{69}{2004}{065009}, \arxiv{\hepth{0309242}}.
							%%CITATION = HEP-TH 0309242;%%

\bibitem{GI-ERG-I}          T.~R.~Morris, \NuclPhys{B}{573}{2000}{97}, \arxiv{\hepth{9910058}}.
							%%CITATION = HEP-TH 9910058;%%


\bibitem{TRM-MassiveScalar} T.~R.~Morris, \NuclPhys{B}{495}{1997}{477}, \arxiv{\hepth{9612117}}.
							%%CITATION = HEP-TH 9612117;%%


\bibitem{Wilson}            K.~Wilson and J.~Kogut, \PhysRep{12}{C}{1974}{75}.
							%%CITATION = PRPLC,12,75;%%


\bibitem{SelfSimilarFlow}   D.~V.~Shirkov, \TheorMathPhys{60}{1985}{778}.
							%%CITATION = TMPHA,60,778;%%

\bibitem{TRM-elements}      T.~R.~Morris, \emph{Prog.\ Theor.\ Phys.\ Suppl.\ } 131 (1998) 395, \arxiv{\hepth{9802039}}.
							%%CITATION = HEP-TH 9802039;%%

\bibitem{scalar1loop}       S.~Arnone, A.~Gatti and T.~R.~Morris \jhep{0205}{2002}{059}, \arxiv{\hepth{0201237}}.
							%%CITATION = HEP-TH 0201237;%%


\bibitem{GeneralMethods}    O.~J.~Rosten, in preparation.


\bibitem{GI-ERG-II}         T.~R.~Morris, \jhep{0012}{2000}{012}, \arxiv{\hepth{0006064}}.
							%%CITATION = HEP-TH 0006064;%%

\bibitem{furry}             H.~Furry, \PhysRev{}{51}{1937}{125}.


\bibitem{itz}               C.~Itzykson and J.~B.~Zuber, `Quantum Field Theory' (1985)
                            McGraw-Hill, Singapore.



\end{thebibliography}
\end{document}